\documentclass[aps,prd, 10pt, twocolumn,nofootinbib,showpacs]{revtex4}
\usepackage{subeqn}
\usepackage[usenames,dvipsnames]{xcolor}

\begin{document}

\title{Self energy in Kerr-Sen space-time incorporated to overspinning and overcharging problems}
\author{Koray D\"{u}zta\c{s}}
\email{koray.duztas@okan.edu.tr}
\affiliation{\.{I}stanbul Okan University, Faculty of Engineering and Natural Sciences, Tuzla, \.{I}stanbul, 34959, T\"{u}rkiye }

\begin{abstract}
Previous attempts to overspin or overcharge Kerr-Sen black holes have been incomplete, as they neglect back-reaction effects. In this study, we incorporate the self-energy of perturbations, resulting from the induced increase in the angular velocity and the electrostatic potential of the event horizon, as described in a seminal work by Will. Our analysis demonstrates that Kerr-Sen black holes cannot be overspun or overcharged by perturbations that satisfy the null energy condition, within a comprehensive second-order framework that includes the self-energy of the perturbations.
\end{abstract} 
\maketitle
\section{Introduction}
The advent of the singularity theorems developed by Penrose and Hawking disrupted the deterministic nature of general relativity \cite{pensing,penhawk}. These theorems prove the geodesic incompleteness of the space-time which is interpreted as the existence of singularities. Geodesic incompleteness implies that photons or freely moving particles may disappear off the edge of the universe. This would render the initial conditions on a Cauchy surface undefined, as the surface intersects the singularity. In that respect, general relativity predicts its own demise. The cosmic censorship conjecture proposed by Penrose provides a way to circumvent this problem \cite{ccc}. In its weak form (wCCC), the conjecture asserts that the singularities that ensue as a result of gravitational collapse, are always shrouded behind the event horizons which disable their causal contact with the observers at the asymptotically flat infinity. This allows the smooth structure of the space-time to be preserved at least outside the black hole region. 

A concrete proof that the gravitational collapse always ends up as a black hole surrounded by an event horizon has been elusive for decades. In the absence of a concrete proof, Wald constructed a thought experiment adapting a new approach. In this type of thought experiments, one starts with a black hole surrounded by an event horizon, then perturbs the black hole with test particles or fields to check the possibility to destroy the event horizon. In the original thought experiment Wald showed that extremal Kerr-Newman black holes cannot be overcharged or overspun into naked singularities \cite{wald74}. Following the seminal work by Wald, many authors constructed similar thought experiments to check the validity of wCCC perturbing the black holes with test particles and fields \cite{hu,js,f1,gao,magne,yuwen,higher,v1,he,wang,jamil,shay3,shay4,zeng,semiz,q1,q2,q3,q4,q5,q6,q7,overspin,emccc,natario,duztas2,mode,coulomb,taubnut,hong,yang,bai,fairoos,tjphys,kerrmog,vasquez,ong}.
Though the notion of a distant observer is not well-defined in the asymptotically de-Sitter and anti-de Sitter cases, the possibility to destroy the event horizon maintains its status as an intriguing problem. Therefore, Wald type tests of wCCC has been extended to these cases \cite{btz,gwak3,chen,ongyao,ghosh,mtz,ext1,he2,dilat,yin,btz1,gwak4,corelli,sia2}. Apparently, generic destructions of event horizons do not occur in the interactions of black holes with test particles or fields satisfying the null energy condition. If the null energy condition is satisfied, there exists a lower bound for the energy of the perturbation to allow its absorption. The contribution of the test particle or field to the mass parameter cannot be less than the lower bound
. This prevents the angular momentum or charge parameters of the space-time to increase beyond the extremal limit. Though non-generic violations of wCCC to second order may occur, these problems are fixed by employing back-reaction effects. An analogous lower limit does not exist for fermionic fields the energy-momentum tensor of which does not satisfy the null energy condition. The low energy modes are also absorbed by black holes, which contribute to the angular momentum and charge parameters, more than the mass parameter. This leads to generic destructions of event horizons into naked singularities \cite{duztas,toth,generic,spinhalf,threehalves}. In this work we restrict ourselves to perturbations satisfying the null energy condition. 

Kerr-Sen black hole is an asymptotically flat solution in the low energy limit of heterotic string theory \cite{sen}. Being a charged, rotating black hole solution it naively appears to have similar aspects with Kerr-Newman black holes. Perturbations and thermodynamics of Kerr-Sen black holes have been extensively studied in the literature \cite{sen1,sen2,sen3,sen4}. In this respect it is also plausible to employ tests of wCCC on Kerr-Sen black holes. In the first  of these thought experiments,  Siahaan showed that charged particles can drive extremal and nearly extremal Kerr-Sen black holes into naked singularities \cite{siahaan}. We also constructed a thought experiments to evaluate the interaction of Kerr-Sen black holes with test fields. We derived that nearly extremal Kerr-Sen black holes can be overspun by test fields with frequencies slightly above the super-radiance limit \cite{kerrsen}. The violations of wCCC derived in these works occur to second order, which suggests that they can be fixed by employing back-reaction effects. Backreaction effects contribute to the interaction to second order and it is anticipated that they fix the overspinning and overcharging problems for non-generic cases.

Backreaction effects have been notoriously difficult to estimate in the interactions of black holes with test particles and fields. Recently, Sorce and Wald derived a condition for the second order perturbations of Kerr-Newman black holes \cite{w2}
\begin{equation}
\delta^2 M-\Omega \delta^2 J-\Phi \delta^2 Q \geq  \frac{\kappa}{8\pi}\delta^2 A
\label{condisorcewald}
\end{equation}
where $\kappa$ is the surface gravity and $A$ is the area of a Kerr-Newman black hole. In a recent paper we evaluated the Sorce-Wald condition (\ref{condisorcewald}) for Kerr and Reissner-Nordstr\"{o}m black holes \cite{spin2}. We showed that the right-hand-side of (\ref{condisorcewald}) depends on $(\delta J)^2$ and $(\delta Q)^2$ respectively, i.e. the back-reaction effects contribute to the interaction to second order as expected. However Sorce and Wald introduce an extra non-physical parameter $\lambda$ and define a function $f(\lambda)$ to determine the final state of the space-time. An order of magnitude analysis for $f(\lambda)$ reveals (see \cite{spin2,absorp})
\begin{equation}
f(\lambda)\sim O(\epsilon^2) - O(\delta M \epsilon \lambda)+ O((\delta M)^2 \lambda^2)
\label{flambda}
\end{equation}
The perturbations $(\delta M)$,$(\delta J)$, $(\delta Q)$ are first order in the test particle/field approximation. In \cite{absorp} we referred to the algebraic fact that $(\delta M)^2 \lambda^2$ is a fourth order quantity, not second. In \cite{spin2} we showed that the Sorce-Wald condition (\ref{condisorcewald}) can be correctly applied to overspinning and overcharging problems by abandoning the non-physical parameter $\lambda$, thereby avoiding the order of magnitude problems. Therefore the Sorce-Wald condition (\ref{condisorcewald}) provides an accurate recipe to calculate the backreaction effects provided that one avoids to multiply their  contribution by the square of an extra non-physical parameter and correctly incorporates their contribution to the analysis, as we have described in \cite{spin2}.

An alternative derivation of back-reaction effects deploys the induced increase in the angular velocity or the electrostatic potential  of the event horizon described in  a seminal work by Will \cite{will}. Will's argument focuses on the interaction of a central rotating mass --which could possibly be charged-- with test particles or fields, rather than the underlying theory from which the black hole solution is derived. Will's method was originally formulated for vacuum solutions such as Kerr and Reissner-Nordstr\"{o}m black holes. The main assumption here is that the course of the interaction is the same for Kerr-Sen black holes which may be justified by the fact that it is an asymptotically flat solution with subtle differences from Kerr-Newman black holes. In particular, analogous thought experiments have been carried out for asymptotically de Sitter and anti-de Sitter space-times, where it is considerably more challenging to envision an interaction proceeding in a similar manner. At $t=0$ test particles or fields are sent in from asymptotically flat infinity to a black hole with specified initial parameters. After a long time $(t \to \infty)$ the perturbations are partially absorbed by the black hole and partially reflected back to infinity, while the parameters of the black hole have been modified due to the initial energy, angular momentum or charge of the perturbations. In addition, Will considers the intermediate stage as the perturbations are incident on the event horizon. The angular momentum and the charge of the perturbations lead to induced increases in the angular velocity and the electrostatic potential of the event horizon, respectively.   The induced increase  also induces a self-energy, which gives an extra contribution to the mass parameter to second order. In \cite{spin2} we showed that the effect of the self-energy derived from Will's argument is effectively equivalent to incorporating the Sorce-Wald condition (\ref{condisorcewald}) into the analysis, for Kerr and Reissner-Nordst\"{o}m black holes. To be more precise the self-energy derived from Will's argument is identically equal to the right-hand -side of (\ref{condisorcewald}). We have employed the induced self-energy in previous works involving both bosonic and fermionic perturbations \cite{kerrmog,spinhalf,threehalves}. Recently, we showed that the induced increase in the angular velocity alone is sufficient to prevent the overspinning of BTZ black holes without any reference to self-energy or gravitational radiation \cite{btzwill}. The main motivation to avoid naked singularities in BTZ space-time, is the fact that they correspond to states rotating faster than light in the AdS/CFT correspondence. Though the overspinning of BTZ black holes is not generic \cite{btz}, the fact that the employment of self-energy or gravitational radiation is problematic in $(2+1)$ dimensions, renders the problem drastic. This drastic problem is solved by considering the fact that the induced increase in the angular velocity of the event horizon, increases the lower limit for the energy of the perturbations to allow their absorption.

In this work we incorporate the self-energy of the perturbations derived from the induced increase in the angular velocity and the electrostatic potential of the event horizon into the overspinning and overcharging problems for Kerr-Sen black holes. The formation of naked singularities in these problems are not generic which suggests that they can be avoided by incorporating the  second order contributions due to self-energy. We examine whether this is indeed the case.

\section{Overspinning problem in Kerr-Sen space-time and self-energy}
Previously, we attempted to overspin Kerr-Sen black holes by scalar test fields \cite{kerrsen}. We derived that there exists a range of frequencies for the test fields that could lead to overspinning of nearly extremal Kerr-Sen black holes. In this section, we review our previous derivation and incorporate the self energy due to the induced increase in the angular velocity of the event horizon. We start with the explicit form of the Kerr-Sen metric in Boyer-Lindquist coordinates:
\begin{widetext}
\begin{equation}
ds^2=-\frac{ \Delta }{ \Sigma }(dt-a\sin^2 \theta d\phi)^2+ \frac{\sin^2 \theta}{\Sigma}[adt-(\Sigma+a^2\sin^2 \theta)d\phi]^2+\Sigma \left( \frac{dr^2}{\Delta }+d\theta^2 \right)
\label{kerrsenmetric}
\end{equation}
\end{widetext}
where $\Delta=r^2-2(M-b)r+a^2$ and $\Sigma=r^2+2br+a^2\cos^2 \theta$. The space-time defined by the metric (\ref{kerrsenmetric}) has three background parameters; namely the mass $M$, the angular momentum $J=Ma$, and charge $Q$ which is implicit in the twist parameter $b=Q^2 /M$.
Kerr-Sen black holes have an inner and an outer horizon located at:
\begin{equation}
r_{\pm}=M-b\ \pm \sqrt{(M-b)^2 -a^2}
\end{equation}
The space-time parameters describe a black hole surrounded by an event horizon if
\begin{equation}
(M-b) \geq a
\label{condi1}
\end{equation}

The quantity $-(g_{t\phi}/(g_{\phi \phi})$ evaluated at the event horizon $r=r_+$ is known as the angular velocity of the event horizon. The observers in the ergosphere --where $g_{tt}$ changes sign-- are compelled to rotate with this angular velocity, to retain their causal time-like character. This effect --known as frame dragging-- occurs when an inertial reference frame is dragged by the rotation of the central mass, which is a characteristic of Kerr-type metrics. For Kerr-Sen space-time the angular velocity of the event horizon can be calculated as:
\begin{equation}
\label{senvelopot:1}
\Omega=\left. -\frac{(g_{t\phi)}}{g_{\phi \phi}}\right|_{r=r_+}=\frac{a}{r_+^2 + 2br_+ +a^2}= \frac{a}{2M r_+} 
\end{equation}

The electrostatic potential of the event horizon is given by:
\begin{equation}
\label{senvelopot:2}
\Phi= \frac{Q}{2M}
\end{equation}
We envisage that a test scalar field incident on the Kerr-Sen black hole from infinity. Our aim is to determine whether the angular momentum parameter of the space-time can be  increased beyond the extremal limit to violate the inequality (\ref{condi1}). For that purpose we should first demand that the test field is absorbed by the black hole. Perturbations satisfying the null energy condition obey Needham's condition to allow their absorption by black holes \cite{needham}.
\begin{equation}
\delta M \geq \Omega \delta J + \Phi \delta Q
\label{needham}
\end{equation}
where $\delta M$, $\delta J$, and $\delta Q$ refer to the perturbations in the mass, angular momentum, and charge parameters of the space-time due to the absorption of the test particle or field. These perturbations are inherently first order quantities in the test particle/field approximation. 

The condition (\ref{needham}) was also independently derived by Natario-Queimada-Vicente \cite{natario} and Sorce-Wald \cite{w2}. However the first derivation known to this author is by Needham. Needham's condition establishes the lower bound for the energy of perturbations relative to their contributions to angular momentum and charge parameters. Perturbations with energies below this threshold are not absorbed by black holes. Allowing the absorption of these modes would result in a generic violation wCCC, as their contributions to angular momentum and charge parameters would be disproportionately large. However, the condition (\ref{condi1}) is not met by fermionic fields, whose energy-momentum tensor does not satisfy the null energy condition. The absorption of low-energy modes for fermionic fields is permitted, posing a significant challenge to wCCC. In this work, we restrict our analysis to perturbations that satisfy the null energy condition.

For  a test field with frequency $\omega$ and azimuthal number $m$, the Needham's condition (\ref{needham}) reduces to the well-known superradiance condition:
\begin{equation}
\omega \geq m\Omega
\end{equation} 
where we have substituted $\delta J=(m/\omega)\delta M$. In \cite{kerrsen} we parametrized nearly extremal Kerr-Sen black holes as:
\begin{equation}
M-b-a=M\epsilon^2 \rightarrow I_{\rm{in}}=2M^2-Q^2-2J=2M^2\epsilon^2
\label{param1}
\end{equation}
Here, we define the indicator function $I(M,J,Q)$, which initially has a value of $2M^2\epsilon^2$. The space-time parameters $(M,J,Q)$ describe a black hole surrounded by an event horizon if $I(M,J,Q) \geq 0$. Equality in this condition corresponds to extremal black holes in the limit $\epsilon \to 0$. If the indicator function becomes negative after the interaction $(I_{\rm{fin}} < 0)$, a real solution for the spatial location of the event horizon $(r_+)$ cannot be found. In such a case, we conclude that the space-time parameters describe a naked singularity. 
We re-write the expression for $r_+$ imposing the parametrization (\ref{param1}):
\begin{equation}
r_+=a+M\epsilon^2+\sqrt{(M-b-a)(M-b+a)}
\end{equation}
In \cite{kerrsen} we have defined the dimensionless parameters $M\alpha^2 \equiv (M-b+a)$ and $\beta^2 \equiv \alpha^2 \epsilon^2$. Notice that $\alpha$ is not necessarily a small parameter. It attains its maximum value as $a \to M $ which equals $\alpha_{\rm{max}} \lesssim 2$. As $a \to 0$, $\alpha$ approaches $\epsilon$. In other words $\alpha$ is strictly larger than $\epsilon$. $\beta$ is larger than $\epsilon$ for $\alpha >1$. In terms of these dimensionless parameters, the spatial location of the event horizon can be written as:
\begin{equation}
r_+=a+M\epsilon^2+M\beta
\end{equation}
The lower bound for the frequency of the test field which corresponds to the superradiance limit, can be expressed as:
\begin{equation}
\omega \geq \omega_{\rm{sl}}=\frac{ma}{2Mr_+}=\frac{m}{2M} \left( \frac{1}{1+\left( \frac{M}{a}\right) \epsilon^2 +\left( \frac{M}{a}\right) \beta} \right)
\label{omegalower}
\end{equation}
The modes with a frequency lower than $ \omega_{\rm{sl}}$ will not be absorbed by the black hole. Next we demand that the black hole is overspun into a naked singularity at the end of the interaction. The final value of the indicator function takes the form:
\begin{equation}
I_{\rm{fin}} = 2(M+ \delta M)^2 - Q^2-2(J+\delta J)
\label{deltafin1}
\end{equation}
If $I_{\rm{fin}}$ can be made negative at the end of the interaction we conclude that the event horizon is destroyed. We substitute $\delta M=M\epsilon$ for the incoming field in accord with the test field approximation and use $\delta J=(m/\omega)\delta M$ to derive the upper limit for the frequency $\omega$ which makes $I_{\rm{fin}}$ negative. One derives that (see \cite{kerrsen})
\begin{equation}
\omega < \omega_{\rm{max}}=\frac{m}{2M(1+\epsilon)}
\label{omegaupper}
\end{equation}
The test field must simultaneously satisfy the two conditions (\ref{omegalower}) and (\ref{omegaupper}) to ensure its absorption by the black hole and the occurrence of overspinning. Therefore, the frequency must lie within the range $\omega_{\rm{sl}} < \omega < \omega_{\rm{max}}$ for overspinning to occur. Additionally, $\beta$ must be greater than $\epsilon$ to ensure that $\omega_{\rm{sl}} < \omega_{\rm{max}}$, which implies that $\alpha > 1$. This imposes a restriction on the angular momentum parameter of the black hole. Note that the definition of $\alpha$ implies:
\begin{equation}
a=\frac{M}{2}(\alpha^2 - \epsilon^2)
\end{equation}
Therefore overspinning occurs for Kerr-sen black holes which initially satisfy
\begin{equation}
a>\frac{M}{2}(1 - \epsilon^2)
\end{equation}
We have not alluded to this fact in \cite{kerrsen}. Let us calculate the minimum value for $I_{\rm{fin}}$ by substituting $\omega=m\Omega$ for the frequency of the incoming field. This implies
\begin{equation}
\delta J=\frac{M\epsilon}{\Omega}=2M^2 \epsilon \left( 1+\left( \frac{M}{a}\right) \epsilon^2 +\left( \frac{M}{a}\right) \beta \right)
\end{equation}
We calculate the minimum value for $I_{\rm{fin}}$ given in (\ref{deltafin1}) by imposing (\ref{param1}), and letting $\omega=m\Omega$:
\begin{equation}
I_{\rm{fin}} \geq 4M^2 \left(\epsilon^2 - \frac{M}{a}\epsilon \beta \right)
\label{deltafinmin}
\end{equation}
The minimum value of $I_{\rm{fin}}$ is negative for $\beta > \epsilon$. However its absolute value is of the order $\epsilon^2$ which suggests that it can be made positive by incorporating the second order contribution from self-energy.  In \cite{kerrsen}, we derived (\ref{omegalower}) and (\ref{omegaupper}) for nearly extremal Kerr-Sen black holes and provided a numerical example where $I_{\rm{fin}}$ becomes negative if self-energy is ignored. We then considered fermionic fields that do not satisfy the null energy condition and, therefore, do not obey Needham's condition. The lower limit (\ref{omegalower}) does not exist for fermionic fields. We presented a numerical example showing that the absorption of a test field with frequency $\omega \sim (m\Omega)/2$ leads to the destruction of the event horizon, as $I_{\rm{fin}} \gtrsim -\epsilon M^2$. The destruction of the event horizon by fermionic fields is generic in the sense that it cannot be corrected by self-energy effects, which contribute to second order.

In this work, we restrict ourselves to scalar fields satisfying the null energy condition. The analytical expression for $I_{\rm{fin}}$ in (\ref{deltafinmin}) will allow us to incorporate the self energy effects analytically. We proceed with the calculation of  the self-energy due to the induced increase in the angular velocity of the event horizon described by Will \cite{will}.  The magnitude of the induced increase can be derived by considering a black hole in the limit $J=0$ and adding an angular momentum $\delta J$ to this black hole. Then the induced increase is given by the angular velocity of the black hole with angular momentum $\delta J$. For a Kerr-Sen black hole, one derives:
\begin{equation}
\Delta \Omega =\lim_{J \to 0}\frac{\delta J}{2M^2 r_+}=\frac{\delta J}{4M^2 (M-b)}
\label{deltaomega}
\end{equation}
where we have substituted $r_+ \to 2(M-b)$ as $J \to 0$. The induced increase in the angular velocity of the test field induces a self-energy:
\begin{equation}
E_{\rm{self}}=(\Delta \Omega)(\delta J)= \frac{(\delta J)^2}{4M^2 (M-b)}
\label{Eself}
\end{equation}
The self-energy derived in (\ref{Eself}) contributes to the mass parameter of the black hole. We will calculate the self-energy for the minimum value of $I_{\rm{fin}}$, which corresponds to the case $\omega = m\Omega$. Specifically, this mode will not be absorbed by the black hole as the angular velocity of the horizon increases ($\Omega \to \Omega + \Delta \Omega$). The lower limit for $\omega$ will be higher, and the range $\omega_{\rm{sl}} < \omega < \omega_{\rm{max}}$ will narrow. However, if the self-energy resolves the overspinning problem for $\omega = m\Omega$, we can conclude that overspinning is not possible for any larger value within the range $\omega_{\rm{sl}} < \omega < \omega_{\rm{max}}$, since the contribution to the angular momentum parameter of the space-time is inversely proportional to the frequency. Therefore, it will be sufficient to examine the effect of self-energy for $\omega = m\Omega$, which implies $\delta J = (M\epsilon)/\Omega$. Substituting this value in (\ref{Eself}):
\begin{equation}
E_{\rm{self}}= \frac{M^2 \epsilon^2}{\Omega^2 4M^2 (M-b)}
\label{Eself1}
\end{equation}
With $\Omega^2 \lesssim 1/(4M^2)$ one finds
\begin{equation}
E_{\rm{self}} \gtrsim \frac{M^2 \epsilon^2}{(M-b)}
\label{Eself2}
\end{equation}
We can re-calculate the minimum value of $I_{\rm{fin}} $, by considering the contribution of self-energy to the mass parameter.
\begin{equation}
I_{\rm{fin}} > 2(M+ \delta M+ E_{\rm{self}})^2 - Q^2-2(J+\delta J)
\label{deltafin2}
\end{equation}
We can substitute the self-energy (\ref{Eself2}) derived for $\omega=m\Omega$. Note that $I_{\rm{fin}} $ will be larger than the right-hand-side of (\ref{deltafin2}), since the modes with $\omega=m\Omega$ will not be absorbed as the angular velocity of the event horizon increases. With this substitution, we calculate $I_{\rm{fin}} $ to second order:
\begin{equation}
I_{\rm{fin}} > 4M^2 \left[ \left( 1+\frac{M}{M-b} \right) \epsilon^2 - \left( \frac{M}{a}\right) \epsilon \beta \right]
\label{deltafin3}
\end{equation}
Note that $(M-b) \sim a$ for a nearly extremal black hole, and $\beta$ attains its maximum value for $a \to M$, $\alpha \to 2$, which equals $\beta \lesssim \sqrt{2}\epsilon$. The right-hand-side of (\ref{deltafin3}) is positive definite. The incorporation of self energy fixes the overspinning problem for Kerr-Sen black holes.

\section{Overcharging problem and self-energy}
The overcharging problem in Kerr-Sen space-time was first examined by Siahaan \cite{siahaan}. It was demonstrated that Kerr-Sen black holes can be overcharged into naked singularities by test particles with specific parameters. This analysis neglects back-reaction effects, which could potentially resolve the overcharging issue. In this section, we assess the overcharging problem and incorporate the self-energy of perturbations to determine whether it could prevent overcharging. We begin with perturbations of energy $\delta M$ and charge $\delta Q$ incident on a Kerr-Sen black hole from infinity. Initially, we require that the perturbations are absorbed by the black hole. Needham's condition establishes the lower bound for the energy of the perturbations to be absorbed by the black hole. With $\delta J=0$, Needham's condition simplifies to:
\begin{equation}
\delta M \geq \Phi (\delta Q)
\label{needhamcharge}
\end{equation}
where $\Phi=Q/(2M)$ for a Kerr-Sen black hole. Let us envisage that we send in a test particle of energy $\delta M=M\epsilon$. Needham's condition (\ref{needhamcharge}) implies that its charge should satisfy:
\begin{equation}
\delta Q \leq \frac{2M^2}{Q}\epsilon
\label{chargemax}
\end{equation}
to ensure that it is absorbed by the black hole. Let us choose the maximum value for the charge of the perturbation and check whether a Kerr-Sen black hole parametrised as (\ref{param1}) can be overcharged by this perturbation. This mode minimizes the value of $I_{\rm{fin}}$ after the interaction.
\begin{eqnarray}
I_{\rm{fin}}&\geq& 2(M+\delta M)^2 -(Q+\delta Q)^2 -2J \nonumber \\
&\geq& 4M^2 \epsilon^2 \left( 1- \frac{M^2}{Q^2} \right) 
\label{deltamincharge}
\end{eqnarray}
where we have substituted $\delta M=M\epsilon$ and $\delta Q=(2M^2 \epsilon)/Q$. When one ignores self-energy, one finds that the minimum possible value of $I_{\rm{fin}}$ is negative at the end of the interaction which indicates the formation of a naked singularity. However, the destruction of the event horizon does not appear to be generic, since the absolute value of $I_{\rm{fin}}$ is of the order of $\epsilon^2$. At this stage, we incorporate the self-energy derived from Will's argument into the analysis. First we calculate the increase in the electrostatic potential. This increase is equal to the electrostatic potential  the black hole would acquire by absorbing the test particle, in the limit $Q \to 0$. By direct substitution:
\begin{equation}
\Delta \Phi=\frac{\delta Q}{2M}
\end{equation}
The induced increase in the electrostatic potential induces the self-energy:
\begin{equation}
E_{\rm{self}}=(\Delta \Phi) (\delta Q)=\frac{(\delta Q)^2}{2M}
\end{equation}
We calculate the self energy for the most challenging mode with $\delta Q=(2M^2 \epsilon)/Q$.
\begin{equation}
E_{\rm{self}}=\frac{2M^3 \epsilon^2}{Q^2}
\label{selfchargeq}
\end{equation}
The self-energy derived in (\ref{selfchargeq}) contributes to the mass parameter of the space-time to modify the minimum value of $I_{\rm{fin}}$ derived in (\ref{deltamincharge}). To second order:
\begin{eqnarray}
I_{\rm{fin}} &\geq & 2(M+ \delta M+E_{\rm{self}})^2 - (Q+\delta Q)^2 -J \nonumber \\
&\geq & 4M^2 \epsilon^2 \left( 1+ \frac{M^2}{Q^2} \right) 
\label{deltafincharge}
\end{eqnarray}
The incorporation of self-energy into the analysis re-assures that the event horizon is preserved at the end of the interaction. Consequently, Kerr-Sen black holes cannot be overcharged by perturbations satisfying the null energy condition.
\section{Conclusions}
Previous works have demonstrated the possibility of overspinning or overcharging Kerr-Sen black holes, disregarding back-reaction effects \cite{siahaan,kerrsen}. In both instances, the destruction of the event horizon does not appear to be generic, suggesting that the event horizon can be restored by employing back-reaction effects. However, back-reaction effects have been notoriously obscure in black hole space-times. In this study, we incorporated the self-energy described in a seminal work by Will, which is derived from the induced increase in the angular velocity or the electrostatic potential of the event horizon \cite{will}. We calculated the self-energy for the most challenging modes, as determined by Needham's condition. Our findings indicate that the magnitude of self-energy is sufficiently large to prevent the overspinning and overcharging of Kerr-Sen black holes. The induced increase in the angular velocity and the electrostatic potential of the event horizon shifts the lower bound for the energy of the perturbations, allowing their absorption by the black hole. Consequently, the most challenging modes would not be absorbed by the black hole. However, this consideration was unnecessary, as the magnitude of the self-energy is sufficiently large to prevent overspinning and overcharging, even for the most challenging modes. We conclude that Kerr-Sen black holes cannot be overcharged or overspun by perturbations satisfying the null energy condition, in a complete second-order analysis that includes the self-energy of the perturbations.


\begin{thebibliography}{99}
\bibitem{pensing} R. Penrose, Phys. Rev. Lett. \textbf{14}, (1965) 57.

\bibitem{penhawk} S.W. Hawking and R. Penrose, Proc. R. Soc. London \textbf{314}, (1970) 529.

\bibitem{ccc} R. Penrose, Rivista del Nuovo Cim. Numero specialle \textbf{1},  (1969) 252.

\bibitem{wald74}
	R.M. Wald, Ann. Phys. (N.Y.) \textbf{82},  (1974) 548.

\bibitem{hu} V.E. Hubeny, Phys. Rev. D \textbf{59}, (1999) 064013.

\bibitem{js}
	T. Jacobson, T.P.Sotiriou,  Phys. Rev. Lett. \textbf{103},   (2009) 141101.
	
\bibitem{f1}  F. de Felice, Y. Yunqiang,  Class. Quantum Grav. \textbf{18},  (2001) 1235.

\bibitem{gao} S. Gao, Y. Zhang, Phys. Rev. D  \textbf{87},   (2013) 044028.	 

\bibitem{magne}H.M.  Siahaan,  Phys. Rev. D \textbf{96},  (2017) 024016.

\bibitem{yuwen} T.Y. Yu, W.Y. Wen,   Phys. Lett. B   \textbf{781}, (2018) 713 .

\bibitem{higher} B. Wu, W. Liu, H. Tang, R.H. Yue,  Int. J. Mod. Phys. A  \textbf{21}, (2017) 1750125 .

\bibitem{v1} K.S. Revelar, I. Vega,  Phys. Rev. D \textbf{96}, (2017) 064010.

\bibitem{he} Y.L. He, J. Jiang,  Phys. Rev. D  \textbf{100}, (2019) 124060 .

\bibitem{wang} P. Wang, H. Wu, H. Yang,  Eur. Phys. J. C  \textbf{79},  (2019) 572.

\bibitem{jamil}K. D\"{u}zta\c{s}, M. Jamil,  Mod. Phys. Lett. A  \textbf{34}, (2019) 1950248 .

\bibitem{shay3} S. Shaymatov, N. Dadhich, B. Ahmedov, M.Jamil,  Eur. Phys. J. C  \textbf{80},  (2020) 481.

\bibitem{shay4} S. Shaymatov, N. Dadhich, Phys. Dark Universe \textbf{31}, (2021) 100758.

\bibitem{zeng} D. Chen, S. Zeng,  Nucl. Phys. B \textbf{957},  (2020) 115089.

\bibitem{semiz} \.{I}.  Semiz, Gen. Relativ. Gravit. \textbf{43}, (2011) 833.

\bibitem{q1}
	G.E.A. Matsas, A.R.R. da Silva, Phys. Rev. Lett.  \textbf{99}, (2007) 181301 .

\bibitem{q2}M. Richartz and A. Saa, Phys. Rev. D  \textbf{78},  (2008) 081503.
	
\bibitem{q3} S. Hod,  Phys. Rev. Lett.  \textbf{100}, (2008) 121101.

\bibitem{q4} S. Hod,  Phys. Rev. D  \textbf{66}, (2002) 024016.

\bibitem{q5} M. Richartz, A. Saa,  Phys. Rev. D  \textbf{84}, (2011)  104021.
	
\bibitem{q6} S. Hod,  Phys. Lett. B  \textbf{668},  (2008) 346.

\bibitem{q7} \.{I}.  Semiz   and   K. D\"{u}zta\c{s},  Phys. Rev. D   \textbf{92}, (2015) 104021.

\bibitem{overspin} K. D\"{u}zta\c{s}  and \.{I}   Semiz,  Phys. Rev. D   \textbf{88}, (2013) 064043. 

\bibitem{emccc}K. D\"{u}zta\c{s}, Gen. Relativ. Gravit. \textbf{46},  (2014) 1709.

\bibitem{natario}J. Natario, L. Queimada, R. Vicente, Class. Quantum Grav. \textbf{33},  (2016) 175002.

\bibitem{duztas2}K. D\"{u}zta\c{s}, \.{I}   Semiz,  Gen. Relativ. Gravit.  \textbf{48},  (2016) 69.

\bibitem{mode}K. D\"{u}zta\c{s},  Phys. Rev. D \textbf{94},  (2016) 044025. 

\bibitem{coulomb} K. D\"{u}zta\c{s}, Eur.  Phys. J. Plus \textbf{132},  (2017) 406.

\bibitem{taubnut}K. D\"{u}zta\c{s} Class. Quantum Grav. \textbf{35},  (2018) 045008.

\bibitem{hong} W. Hong, B. Mu, J. Tao, Nucl. Phys. B  \textbf{949},  (2019) 114826.

\bibitem{yang} D. Chen, W. Yang, X. Zeng,  Nucl. Phys. B  \textbf{946},  (2019) 114722.

\bibitem{bai} T. Bai, W. Hong, B. Mu, J. Tao,  Commun. Theor. Phys. \textbf{72},  (2020) 015401.

\bibitem{fairoos} C. Fairoos, A. Ghosh, S. Sarkar, Phys. Rev. D \textbf{96}, (2017) 084013.

\bibitem{tjphys} K. D\"{u}zta\c{s},  Turk. J. Phys. \textbf{42},  (2018) 11.

\bibitem{kerrmog} K. D\"{u}zta\c{s}, Eur. Phys. J. C  \textbf{80},  (2020) 19.

\bibitem{vasquez} I.R. V\'{a}squez, Eur. Phys. J. C  \textbf{84},  (2024) 1197.

\bibitem{ong} Y.C. Ong,  Int. J. Mod. Phys. A  \textbf{35}, (2020) 2030007.

\bibitem{btz}K. D\"{u}zta\c{s},  Phys. Rev. D   \textbf{94},  (2016) 124031.

\bibitem{gwak3} B. Gwak, J. Cosmol. Astropart. Phys.  \textbf{08}, (2019) 016 .

\bibitem{chen} D. Chen,  Eur. Phys. J. C  \textbf{79},  (2019) 353.

\bibitem{ongyao} Y.C. Ong, Y. Yao,  J. High Energy Phys. \textbf{10},  (2019) 129.

\bibitem{ghosh} R. Ghosh, C. Fairoos, S. Sarkar,  Phys. Rev. D   \textbf{100},  (2019) 124019.

\bibitem{mtz}K. D\"{u}zta\c{s}, M. Jamil, S. Shaymatov, B. Ahmedov,  Class. Quantum Grav.  \textbf{37}, (2020) 175005 .

\bibitem{ext1} R. Ghosh, A.K. Mishra, S. Sarkar, Phys. Rev. D  \textbf{104},  (2021) 104043.

\bibitem{he2} K.J. He, G.P. Li, X.Y. Hu,  Eur. Phys. J. C  \textbf{80}, (2020) 209.

\bibitem{dilat}K. D\"{u}zta\c{s}, M. Jamil,  Int. J. Geom. Methods Mod. Phys. \textbf{17}, (2020) 2050207.

\bibitem{yin} R. Yin,  J. Liang, B. Mu,   Phys. Dark Universe  \textbf{32},  (2021) 100831.

\bibitem{btz1} A.K. Ahmed, S. Shaymatov, B. Ahmedov,  Phys. Dark Universe  \textbf{37}, (2022) 101082.

\bibitem{gwak4} B. Gwak,  J. Cosmol. Astropart. Phys.  \textbf{10}, (2021) 012.

\bibitem{corelli} F. Corelli, M. De Amicis, T. Ikeda, P. Pani,  Phys. Rev. D  \textbf{107},  (2023) 044061.

\bibitem{sia2} H.M. Siahaan, P.C. Tijang, Int. J. Mod.  Phys. D  \textbf{32},  (2023) 2250140.

\bibitem{duztas}K. D\"{u}zta\c{s}, Class. Quantum Grav.  \textbf{32}, (2015) 075003 .

\bibitem{toth}G.Z. Toth,  Class. Quantum Grav. \textbf{33},  (2016) 115012.

\bibitem{generic} K. D\"{u}zta\c{s},  Eur. Phys. J. C  \textbf{79}, (2019) 316.

\bibitem{spinhalf} K. D\"{u}zta\c{s},  Eur. Phys. J. C \textbf{81},  (2021) 1131.

\bibitem{threehalves} K. D\"{u}zta\c{s}, Eur. Phys. J. C \textbf{83}, (2023) 567.

\bibitem{sen} A. Sen,  Phys. Rev. Lett. \textbf{69}, (1992) 1006.

\bibitem{sen1} K. Hioki and U. Miyamoto, Phys.
Rev. D \textbf{78},   (2008) 044007.

\bibitem{sen2} A. M. Ghezelbash and H.M. Siahaan, Class.
Quantum Grav. \textbf{30}, (2013) 135005.

\bibitem{sen3} H. M. Siahaan,  Int. J. Mod. Phys.
D \textbf{24}, (2015) 1550102.

\bibitem{sen4} H. L. Prihadi, F. P. Zen, D. Dwiputra, S. Ariwahjoedi,  Phys. Rev. D  \textbf{107}, (2023) 124053.

\bibitem{siahaan} H.M. Siahaan, Phys. Rev. D  \textbf{93},  (2016) 064028.

\bibitem{kerrsen}K. D\"{u}zta\c{s},   Int. J. Mod. Phys. D \textbf{28},  (2019) 1950044.

\bibitem{w2} J. Sorce and R.M. Wald,  Phys. Rev. D \textbf{96},  (2017) 104014.

\bibitem{absorp} K. D\"{u}zta\c{s},  Eur. Phys. J. C  \textbf{81}, (2021) 49.

\bibitem{spin2} K. D\"{u}zta\c{s},  Phys. Rev. D    \textbf{110}, (2024) 024081.

\bibitem{will} C.M. Will, Astrophys. J.  \textbf{191},  (1974) 521.

\bibitem{btzwill} K. D\"{u}zta\c{s},  Eur. Phys. J. C \textbf{84}, (2024) 669.

\bibitem{needham}T. Needham,  Phys. Rev. D \textbf{22}, (1980) 791.
\end{thebibliography}
\end{document}